\documentclass[12pt,english,article,amsmath,amssymb]{article}
\usepackage{graphicx,graphics}
\usepackage{dcolumn}
\usepackage{amsmath,amssymb,amsfonts}
\usepackage{latexsym,verbatim}
\usepackage{bm,bbm}
\usepackage{color}
\usepackage{ulem}
\usepackage[T1]{fontenc}
\usepackage[latin1]{inputenc}
\usepackage{dcolumn}
\usepackage{color}
\usepackage{babel}
\usepackage[breaklinks=true,colorlinks,citecolor=blue,linkcolor=blue,urlcolor=blue]{hyperref}
\usepackage{braket}
\usepackage[percent]{overpic}

\DeclareMathAlphabet\mathbfcal{OMS}{cmsy}{b}{n}
\begin{document}
\title{Quantum supercapacitors}
\author{Dario Ferraro$^{1,2}$\footnote{Corresponding author: ferraro@fisica.unige.it},  Gian Marcello Andolina$^{1,3}$, Michele Campisi$^{4,5}$, \\ Vittorio Pellegrini$^{1,6}$, and Marco Polini$^{1}$}
\date{}
\maketitle
\noindent$^1$ Istituto Italiano di Tecnologia, Graphene Labs, Via Morego 30, I-16163 Genova, Italy. \\
\noindent$^2$ Dipartimento di Fisica, Universit\`a di Genova, Via Dodecaneso 33, I-16146, Genova, Italy.\\
\noindent$^3$ NEST, Scuola Normale Superiore, I-56126 Pisa, Italy. \\
\noindent$^4$ Department of Physics and Astronomy, University of Florence, Via Sansone 1, I-50019, Sesto Fiorentino (FI), Italy.\\
\noindent$^5$ INFN Sezione di Firenze, via G. Sansone 1, I-50019, Sesto Fiorentino (FI), Italy.\\
\noindent$^6$ Bedimensional S.p.a., Via Albisola 121, I-16163 Genova, Italy. \\

\maketitle

{\bf Abstract.} Recently there has been a great deal of interest on the possibility to exploit quantum-mechanical effects to increase the performance of energy storage systems. Here we introduce and solve a model of a {\it quantum supercapacitor}. This consists of two chains, one containing electrons and the other one holes, hosted by arrays of double quantum dots, the latter being a building block of experimental architectures for realizing charge and spin qubits. The two chains are in close proximity and embedded in the same photonic cavity, which is responsible for long-range coupling between all the qubits, in the same spirit of the Dicke model. By employing a variational approach, we find the phase diagram of the model, which displays ferromagnetic and antiferromagnetic phases for suitable pseudospin degrees of freedom, together with phases characterized by collective superradiant behavior. Importantly, we show that when transitioning from the ferro/antiferromagnetic  to the superradiant phase, the quantum capacitance of the model is greatly enhanced. Our work offers opportunities for the experimental realization of a novel class of quantum supercapacitors with an enhanced storing power stemming from exquisite quantum mechanical effects.\\

One of the main challenges of nowadays technology is represented by energy storage~\cite{Crabtree15}. In this context, devices like batteries~\cite{Vincent_book, Scrosati_book} and supercapacitors~\cite{Wang12, Badwal14} are currently employed in a plethora of applications ranging from personal electronics to the automotive sector. Supercapacitors, in particular, are improved versions of conventional capacitors that exploit a molecular-scale interface between the ions of an electrolyte and the electrode to increase the energy density while displaying large power densities. These devices operate on the basis of extremely robust electrical and electrochemical principles developed during the Eighteenth and Nineteenth centuries~\cite{Vincent_book}. However, the progressively growing demand for storage capability and power calls for the elaboration of new strategies. Although a great deal of effort is currently focused on  optimizing materials~\cite{Liu10, Bonaccorso15}, fundamental research in this field may lead, in the long run, to a paradigmatic shift. 

An intriguing possibility is to use quantum resources to boost the charging power density of a battery or the stored energy density of a supercapacitor. Quantum phenomena may indeed enable superior performance of technological devices of various sorts. A seminal example is quantum computing~\cite{DiVincenzo95} performed with quantum bits as compared to classical binary logic. While in this framework quantum mechanics is employed to achieve efficient manipulation and processing of information, an increasing theoretical and experimental research activity is currently focused on applying quantum resources to improve energy storage and transfer~\cite{Campisi11, Horodecki13, Goold16, Vinjanampathy16, Strasberg16, Karimi16, Karimi17}. In particular, a number of researchers have been recently working on {\it{quantum batteries}}~\cite{Alicki13, Hovhannisyan13, Binder15, Campaioli17,  Friis17, Ferraro18, Le18, Andolina18,Andolina18b,Farina18,Julia-Farre18,Zhang18,Andolina18c,Zhang18b,Barra19,Campaioli18}. A solid-state implementation of a quantum battery based on an array of $N$ two-level systems coupled to a common cavity photonic mode (known as Dicke model~\cite{Dicke54, Garraway11, Kirton18}) has been introduced in Ref.~\cite{Ferraro18}. 

Quantum effects, such as exchange and correlations in low-dimensional electron systems with long-range Coulomb interactions~\cite{Giuliani05}, constitute powerful tools that can be potentially manipulated and engineered for energy storage applications. On general grounds, the electronic contribution $C_{\rm e}$ to the capacitance of a mesoscopic device can be written as $C^{-1}_{\rm e}=C^{-1}_{\mathrm{g}}+C^{-1}_{\mathrm{q}}$, where $C_{\mathrm{g}}$ is a classical contribution, i.e.~the conventional geometric capacitance, and $C_{\mathrm{q}}$ is a quantum contribution, usually termed ``quantum capacitance''. The latter accounts for the variation of the Fermi energy due to charge accumulation~\cite{Luryi88,Giuliani05}, i.e.~$C_{\rm q} = Se^2 \partial n/\partial \mu$ where $S$ is the area of the device, $-e$ is the elementary charge, $\mu$ is the chemical potential, and $n$ the electron density. Usually, $C_{\rm q}>0$ and the quantum contribution has therefore the net effect to lower the capacitance of the device, thereby reducing the stored energy density with respect to the classical case, as predicted~\cite{Barlas07,Hwang07} and observed in graphene~\cite{Xia09, Droscher10,Yu13,ji_naturecommun_2014}, for example.

However, situations where a negative exchange and correlation contribution to the energy dominates over the positive kinetic energy do exist. In this case, the compressibility $K = n^{-2}\partial n/\partial \mu$ of the electron gas is negative~\cite{Giuliani05,Kopp09,hroblak_prb_2017} leading to $C_{\rm q}<0$ and $C>C_{\rm g}$. Such quantum mechanical enhancement of the total capacitance as compared to the classical value has been observed in several systems including two-dimensional electron double layers formed in GaAs semiconductor quantum wells~\cite{Eisenstein92}, the interface between two oxides (LaAlO$_{3}$/SrTiO$_{3}$)~\cite{Li11}, 2D monolayers of WSe$_{2}$~\cite{Riley15}, and graphene-MoS$_{2}$ heterostructures~\cite{Larentis14}. 

Negative compressibility is ultimately due to charge rearrangement. In this work we investigate capacitance enhancement effects stemming also from charge rearrangement but this time due to light-matter interactions and ground-state macroscopic quantum coherence. To this end, we focus on two chains of double quantum dots (DQDs)~\cite{van der Wiel02}, one filled with electrons the other with holes, which implement a collection of charge qubits~\cite{Hayashi03, Gorman05, Wang13, Ota18}. All DQDs are coupled to a common photonic cavity mode as e.g.~in Refs.~\cite{Childress04, Frey12, Toida13, Basset13, Stockklauser17}. Each chain separately can be modelled via the Dicke-Ising model~\cite{Zhang13}. The two chains, however, are coupled via an on-site electron-hole attractive interaction, which brings in new qualitative features.  By employing an essentially analytical variational approach, we first demonstrate that this system displays a rich and intriguing ground-state phase diagram as a function of intra- and inter-chain  interactions, and of the coupling between the DQDs and radiation. We then show that the capacitance of the model is dramatically enhanced by light-matter interactions due to the charge rearrangement that occurs at the superradiant phase transition~\cite{Wang73, Hioe73, Emary03}. Finally, we conclude by mentioning that our {\it{quantum supercapacitor}} model is amenable, at least in principle, to be quantum simulated using solid-state architectures~\cite{Singha11, Hensgens17, Reiner16} comprising semiconducting and metallic elements.

{\bf Results}

{\bf Model.}

\begin{figure}[h]
\centering
\begin{overpic}[width=9.2cm]{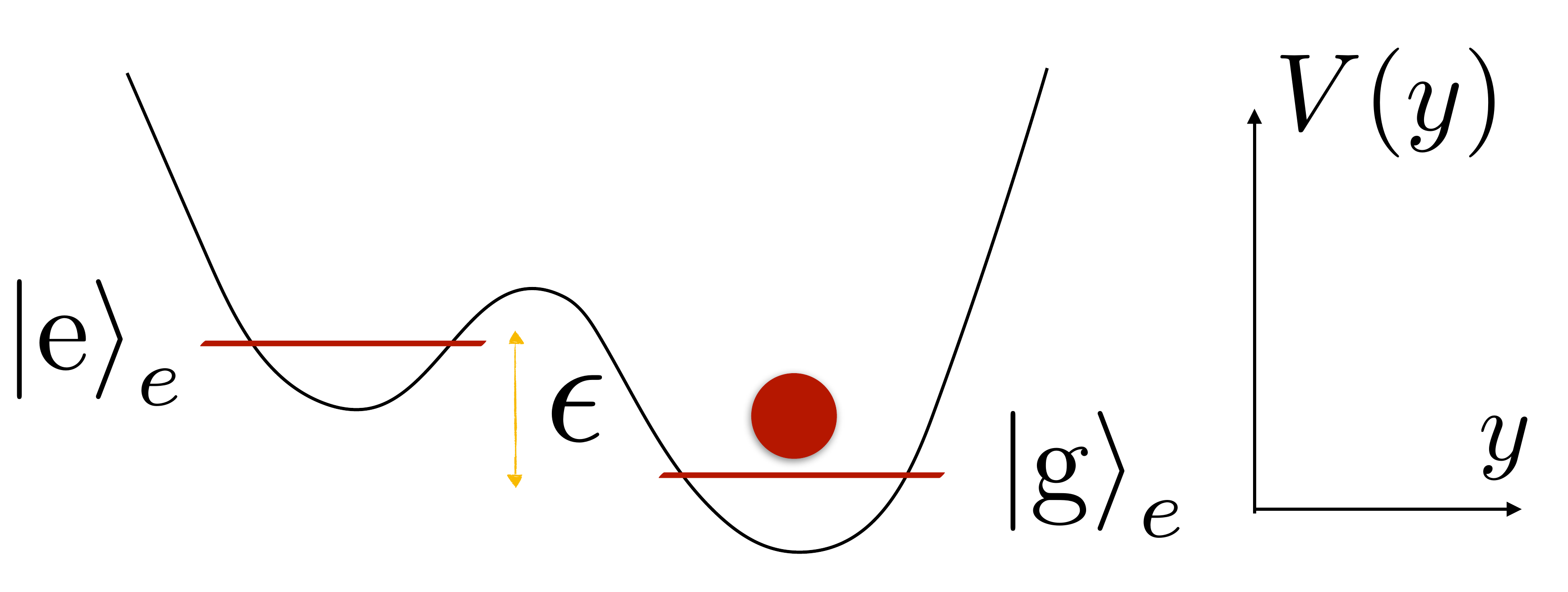}
\put(-4,47){\textbf{a}}
\end{overpic}\\
\vspace{1 cm}
\begin{overpic}[width=9.2cm]{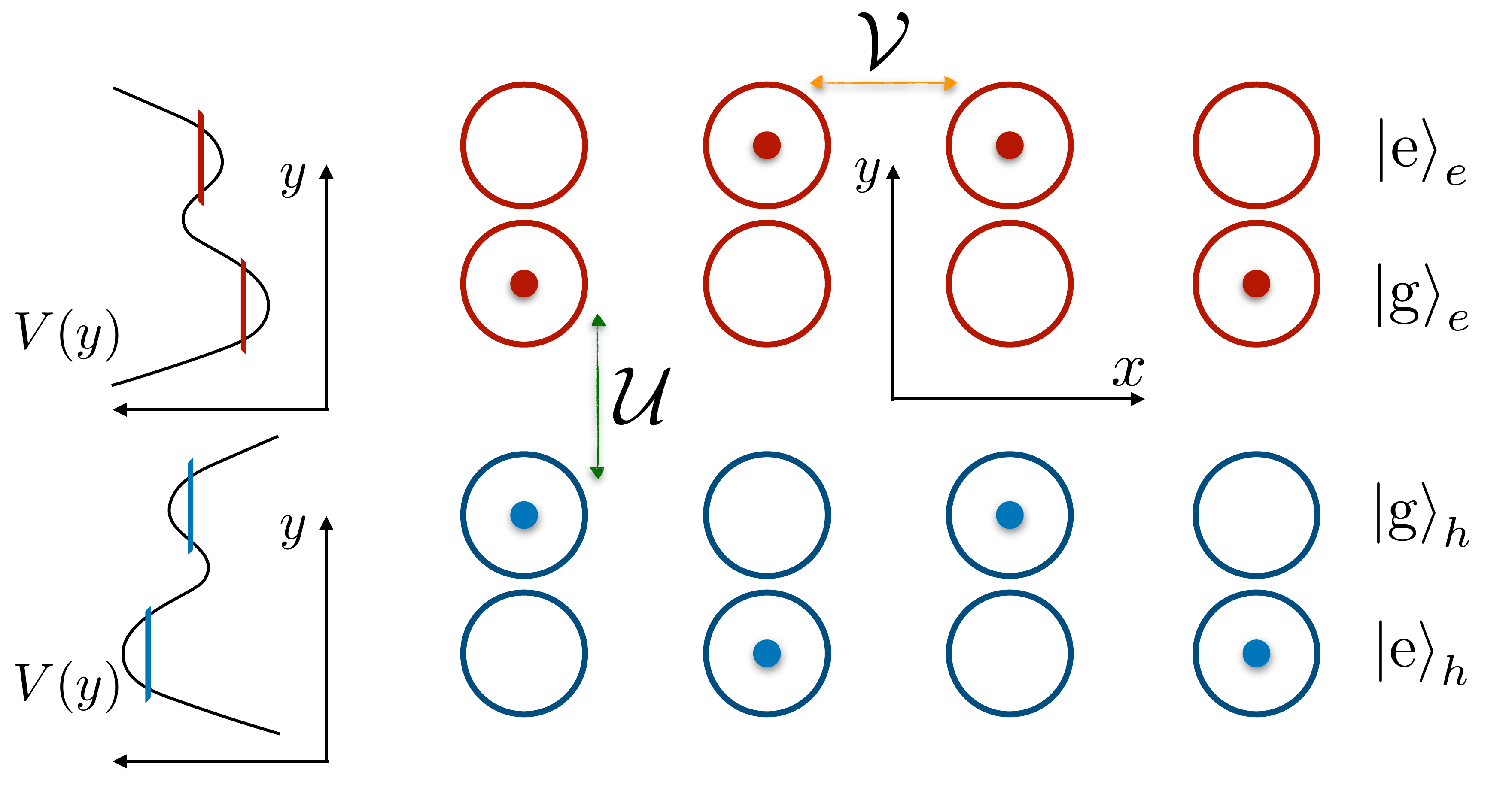}
\put(-4,47){\textbf{b}}
\end{overpic}
\caption{{\textbf{Double quantum dot and two-chain geometry.}} (\textbf{a}) Voltage profile of a double quantum dot occupied by a single electron. The ground ($|{\rm{g}}\rangle_{e}$) and excited ($|{\rm{e}}\rangle_{e}$) states are geometrically separated and the energy gap between them is $\epsilon$. The $|{\rm{g}}\rangle_{e} \rightarrow |{\rm{e}}\rangle_{e}$ transition can be induced by the absorbtion of photons (viceversa the $|{\rm{e}}\rangle_{e} \rightarrow |{\rm{g}}\rangle_{e}$ transition occurs through photon emission). (\textbf{b}) Schematic top view of the two-chain system. Here, one has a top (T) chain (red) and a bottom (B) chain (light blue) made up of double quantum dots. Each of them is singly occupied with electrons (dark red) and holes (dark blue) respectively. Two dominant contributions to the electrostatic interaction have been analyzed in this work: i) an inter-chain attractive interaction of strength $\mathcal{U}$ (green arrow) between an electron and a hole in their respective ground states; and ii) an intra-chain repulsive interaction of strength $\mathcal{V}$ (orange arrow) between electrons (holes) either in the $|{\rm{g}}\rangle_{e}-|{\rm{g}}\rangle_{e}$ ($|{\rm{g}}\rangle_{h}-|{\rm{g}}\rangle_{h}$) or $|{\rm{e}}\rangle_{e}-|{\rm{e}}\rangle_{e}$ ($|{\rm{e}}\rangle_{h}-|{\rm{e}}\rangle_{h}$) configuration.}
\label{fig1}
\end{figure}

Below we investigate arrays of DQDs~\cite{van der Wiel02}, each one with a voltage profile as sketched in Fig.~\ref{fig1}a. Each DQD can be seen as a charge qubit where it is possible to identify a ground ($|{\rm{g}}\rangle$) and an excited ($|{\rm{e}}\rangle$) state, which are separated by an energy gap $\epsilon$ and also spatially~\cite{Hayashi03,Gorman05,Wang13,Ota18}. 

We consider two coupled chains, each one containing $N$ DQDs. The chemical potential in the top chain (T) is tuned in such a way to host exactly one electron in each DQD, while in the bottom one (B) each DQD is filled with one hole (see Fig.~\ref{fig1}b). Such charge configuration has been chosen in order to mimic the two oppositely charged plates of a classical capacitor. In our toy model we assume that only two (screened) Coulomb interaction terms are at play: i) an inter-chain {\it attractive} interaction of strength ${\cal U}$ between an electron in state $|{\rm{g}}\rangle_{e}$ on site $i$ of the T chain and a hole in state $|{\rm{g}}\rangle_{h}$ on the corresponding site of the B chain and ii) an intra-chain {\it repulsive} interaction of strength ${\cal V}$ between two electrons (holes) in states $|{\rm{g}}\rangle_{e}$ ($|{\rm{g}}\rangle_{h}$) or $|{\rm{e}}\rangle_{e}$ ($|{\rm{e}}\rangle_{h}$) in the T (B) chain, acting only between adjacent DQDs. Finally, in each DQD, the transition between the ground state (excited state) and the excited state (ground state) is induced by absorption (emission) of photons from (into) the electromagnetic field of a cavity. We model this interaction via a Dicke-type coupling~\cite{Dicke54,  Garraway11, Kirton18} between the cavity photonic mode and the DQDs, effectively behaving as two-level systems.

In this framework, our model is  described by the following Hamiltonian (see Methods section):
\begin{equation}\label{Starting_H}
\hat{\mathcal{H}} = \hat{\mathcal{H}}^{({\rm T})}_{\rm DI} + \hat{\mathcal{H}}^{({\rm B})}_{\rm DI} + \hat{\mathcal{H}}^{({\rm TB})} + \hat{\mathcal H}^{({\rm R})}~,
\end{equation}
where the first (second) term describes the T (B) chain of DQDs and their interactions with the cavity mode, 
\begin{equation}\label{eq:DIT}
\hat{\mathcal{H}}^{({\rm T})}_{\rm DI}=\frac{\epsilon}{2} \sum_{i=1}^{N} \hat{\tau}^{z}_{i}+\frac{\mathcal{V}}{2} \sum_{i=1}^{N} \left( \hat{\tau}^{z}_{i} \hat{\tau}^{z}_{i+1}+1\right) + \hbar \omega_{\mathrm c} \lambda \left(\hat{a}^{\dagger}+\hat{a} \right)\sum_{i=1}^{N} \hat{\tau}^{x}_{i}~,
\end{equation} 
the third term describes inter-chain local attractive interactions (${\cal U}>0$),
\begin{equation}\label{eq:intra_chain}
\hat{\mathcal{H}}^{({\rm TB})} = -\frac{\mathcal{U}}{4} \sum_{i=1}^{N} \left(1- \hat{\tau}^{z}_{i} \right)\left(1- \hat{\sigma}^{z}_{i} \right)~,
\end{equation}
and the fourth term describes the cavity radiation mode,
\begin{equation}\label{eq:radiation}
\hat{\mathcal{H}}^{({\rm R})} = \hbar \omega_{\mathrm c} \hat{a}^{\dagger}\hat{a}~.
\end{equation}
The B chain Hamiltonian, $\hat{\mathcal{H}}^{({\rm B})}_{\rm DI}$, can be obtained from $\hat{\mathcal{H}}^{({\rm T})}_{\rm DI}$ by replacing $\hat{\tau}^{\alpha}_{i} \to \hat{\sigma}^{\alpha}_{i}$, where $\hat{\tau}^{\alpha}_{i}$ ($\hat{\sigma}^{\alpha}_{i}$) with $\alpha = x, z$ are pseudospin Pauli matrices acting on the 2D Hilbert space associated with the $i$-th DQD on the T (B) chain. In Eqs.~(\ref{eq:DIT}) and~(\ref{eq:radiation}), $\hat{a}$ ($\hat{a}^{\dagger}$) is the annihilation (creation) operator for a cavity photon of frequency $\omega_{\mathrm{c}}$, and $\lambda$ is a dimensionless parameter describing the strength of the coupling between cavity photons and each DQD~\cite{Dicke54, Garraway11, Kirton18}. 

It is worth noticing that the pseudospin part of the above Hamiltonian is a multi-DQDs generalization of the model discussed in Refs.~\cite{Petersson09, Li15, Ward16} where CNOT gates for two capacitively coupled charge qubits were investigated. Moreover, Eq.~(\ref{Starting_H}) can be seen as two copies of the Dicke-Ising (DI) model introduced in Ref.~\cite{Zhang13}, one for the T chain described by $\hat{\mathcal{H}}^{({\rm T})}_{\rm DI}$ and one for the B chain described by $\hat{\mathcal{H}}^{({\rm B})}_{\rm DI}$, further coupled by means of the local attractive interaction $\propto \mathcal{U}$. Since the DI model shows a non-trivial phase diagram in the $\mathcal{V}$-$\lambda$ space, this is inherited by our model. In addition, however, we expect more ground-state phases triggered by ${\cal U}>0$. In particular, we expect a ferromagnetic arrangement with electrons and holes all in the ground state for large values of $\mathcal{U}$ and an antiferromagnetic ordering with electrons and holes alternatively in the $|{\rm{g}}\rangle$ and the $|{\rm{e}}\rangle$ state for large values of $\mathcal{V}$. These are expected to coexist with an overall normal/superradiant phase transition~\cite{Wang73, Hioe73, Emary03} driven by $\lambda$. The interplay between these competing phases leads to an extremely rich phenomenology that will be investigated in the following by means of a variational technique consisting in classifying the stable phases of the system's ground state.

{\bf Ground-state energy and quantum phase transitions.} 

We begin by writing down the following variational ground-state wave-function for the problem at hand: 
\begin{eqnarray}\label{wavefunction}
|\Psi\rangle =|\sqrt{N} \alpha \rangle \otimes \prod_{i=1}^{N}
\left( \begin{array}{cc}
\cos \left(\frac{\theta^{({\rm{B}})}_{i}}{2}\right)\\
e^{i \chi^{({\rm{B}})}_{i}}\sin \left(\frac{\theta^{({\rm{B}})}_{i}}{2}\right)
\end{array}\right)
\otimes
\prod_{k=1}^{N}
\left( \begin{array}{cc}
\cos \left(\frac{\theta^{({\rm {T}})}_{k}}{2}\right)\\
e^{i \chi^{({\rm{ T}})}_{k}}\sin \left(\frac{\theta^{( {\rm{T}})}_{k}}{2}\right)
\end{array}\right)~.
\end{eqnarray}
Here, $|\sqrt{N} \alpha \rangle$ denotes a coherent state of the cavity with displacement $\sqrt{N}\alpha$ (assumed real for the sake of simplicity)~\cite{Emary03} and $\theta^{({\rm{T/B}})}_{i}, \chi^{({\rm{T/B}})}_{i}$ the angles characterizing the Bloch state of the pseudospin associated with the $i$-th DQD in the T or B chain, respectively. 
Note that $\theta^{({\rm{T/B}})}_{i}\neq 0, \pi$ denotes states which are coherent quantum superpositions of $|{\rm{g}}\rangle$ and $|{\rm{e}}\rangle$.
We consider periodic boundary conditions $(N+1\equiv 1)$, as usually done in the study of Heisenberg spin chains~\cite{Baxter82}, which imply translational invariance of the completely filled chains. Moreover, we exploit the ${\rm{T}}\leftrightarrow {\rm{B}}$ exchange symmetry of the model, which allows us to set $\theta^{({\rm{T}})}_{i}=\theta^{({\rm{B}})}_{i}= \theta_{i}$ and $\chi^{({\rm{T}})}_{i}=\chi^{({\rm{B}})}_{i}= \chi_{i}$. Accordingly, the ground-state energy $E=\langle \Psi| \hat{\mathcal{H}} | \Psi \rangle$ of the completely filled system is given by
\begin{eqnarray}
E&=& \sum_{i=1}^N \Bigg[\left({\epsilon}+\frac{\mathcal{U}}{2} \right) \left(\cos\theta_{i} \right)+\mathcal{V} \left(\cos\theta_{i} \cos\theta_{i+1} \right) - 
\frac{\mathcal{U}}{4} \left(\cos^{2}\theta_{i}\right) \nonumber\\
&+& 4\hbar \omega_{\mathrm{c}} \lambda \sqrt{N} \alpha \left(\sin\theta_{i} \cos\chi_{i}\right) + \hbar \omega_{\mathrm{c}} \alpha^{2} +\mathcal{V}- \frac{\mathcal{U}}{4}\Bigg]~.
\end{eqnarray}
Assuming $N$ to be even, and restricting the analysis to the case in which the polar $\theta_{i}$ and azimuthal $\chi_{i}$ angles can only change between even and odd sites~\cite{Zhang13}, i.e.~$ \theta_{2i+1}=\theta_{\rm o}$, $\theta_{2i}=\theta_{\rm e}$, $\chi_{2i+1}=\chi_{\rm o}$, $\chi_{2i}=\chi_{\rm e}$, we finally obtain
\begin{eqnarray}\label{functional}
E&=& N \left[\left(\frac{\epsilon}{2}+\frac{\mathcal{U}}{4} \right) \left(\cos\theta_{\rm o}+\cos\theta_{\rm e} \right)+\mathcal{V} \left(\cos\theta_{\rm o} \cos\theta_{\rm e} \right) \right. \nonumber \\ 
&-&\left.\frac{\mathcal{U}}{8} \left(\cos^{2}\theta_{\rm o}+\cos^{2}\theta_{\rm e} \right)+ 2\hbar \omega_{\mathrm{c}} \lambda \sqrt{N} \alpha \left(\sin\theta_{\rm o} \cos\chi_{\rm o}+\sin\theta_{\rm e} \cos\chi_{\rm e}\right) \right. \nonumber \\ 
&+&\left.\hbar \omega_{\mathrm{c}} \alpha^{2} +\mathcal{V}- \frac{\mathcal{U}}{4}\right]~.
\end{eqnarray}
The function $E = E(\theta_{\rm o},\theta_{\rm e}, \chi_{\rm o}, \chi_{\rm e}, \alpha)$ needs to be minimized with respect to its 5 variables in order to identify the ground-state energy of the system, to characterize possible stable phases, and transitions between them. Minimizing with respect to $\chi_{\rm o} $ and $\chi_{\rm e}$  we get
\begin{eqnarray}
\frac{\partial E}{\partial\chi_{\rm e}}&=&- 2\hbar \omega_{\mathrm{c}} \lambda \sqrt{N} \alpha \left(\sin\theta_{\rm e} \sin\chi_{\rm e} \right)=0~,\\
\frac{\partial E}{\partial\chi_{\rm o}}&=&- 2\hbar \omega_{\mathrm{c}} \lambda \sqrt{N} \alpha \left(\sin\theta_{\rm o} \sin\chi_{\rm o}\right)=0~,
\end{eqnarray}
which, excluding $\alpha=0$ and  $\sin\theta_{\rm e} = \sin\theta_{\rm o} = 0$, are solved by $\chi_{\rm e}=l \pi$ and $\chi_{\rm o}=m \pi$, with $l,m$ integers. Hence, $\cos{\chi_{\rm o}}$ and $\cos{\chi_{\rm e}}$ in Eq.~(\ref{functional}) reduce to an overall $\pm $ sign, which can be seen as a redefinition of $\theta_{\rm o}$ and $\theta_{\rm e}$.  Therefore, in the framework of our variational approach, we can fix $\cos{\chi_{\rm o}}=\cos{\chi_{\rm e}}=1$.

To further reduce the number of variables involved in the problem, it is convenient to minimize first with respect to $\alpha$, i.e.~$\partial E/\partial \alpha=0$, which yields
\begin{equation}
\alpha=- \lambda \sqrt{N}\left(\sin \theta_{\rm o} +\sin \theta_{\rm e} \right)~,
\end{equation}
and replace the latter result into Eq.~(\ref{functional}). From the previous equation we notice that, if superradiance occurs, i.e.~if $\alpha\neq0$, then $ \theta_{\rm o}, \theta_{\rm e}\neq 0, \pi$, which, for the reasons stated above after Eq.~(\ref{wavefunction}), implies that coherence occurs between $|{\rm{g}}\rangle_{e}$ and  $|{\rm{e}}\rangle_{e}$. Superradiance is related with the emergence of macroscopic coherence in the ground-state wave-function~\cite{Emary03}.

This leads to the simplified ground-state energy function
\begin{equation}
\tilde{E}=N \left[\left(\epsilon+\frac{\mathcal{U}}{2}\right) s +\mathcal{V} \left(s^{2}-m^{2}\right)- \frac{\mathcal{U}}{4}  \left(s^{2}+m^{2}\right)-\hbar \omega_{\mathrm{c}}\mathcal{A} +\mathcal{V}- \frac{\mathcal{U}}{4}\right]~,
\end{equation}
written in terms of the order parameters~\cite{Zhang13}
\begin{eqnarray}\label{eq:order_parameters}
\mathcal{A} &=&\frac{\langle \hat{a}^{\dagger} \hat{a} \rangle}{N} =\Lambda^{2} \left(\sin\theta_{\rm o}+\sin\theta_{\rm e} \right)^{2}\nonumber\\
s&=&\frac{\langle \hat{\sigma}^{z}_{1}+\hat{\sigma}^{z}_{2} \rangle}{2}= \frac{1}{2} \left(\cos \theta_{\rm o}+\cos \theta_{\rm e} \right)\nonumber\\
m&=&\frac{\langle \hat{\sigma}^{z}_{1}-\hat{\sigma}^{z}_{2} \rangle}{2}= \frac{1}{2} \left(\cos \theta_{\rm o}-\cos \theta_{\rm e} \right)~.
\end{eqnarray}
The physical interpretation of these quantities is the following. The first one, ${\cal A}$, measures the average number of photons in the cavity and is non-zero in the superradiant phase. The second one, $s$, is the magnetization of a plaquette composed of two neighboring sites. We stress that $s^2\neq 1$ implies macroscopic coherence in the many-particle wave-function. Finally, $m$ is the plaquette staggered magnetization. The quantity $\tilde{E}$ has been minimized numerically with respect to $\theta_{\rm o}$ and $\theta_{\rm e}$, as a function of the dimensionless parameters $\tilde{\mathcal{U}}  \equiv \mathcal{U}/\hbar \omega_{\mathrm{c}}$ and $\Lambda^{2}\equiv \lambda^{2}N$ and for different values of $\tilde{\mathcal{V}}\equiv \mathcal{V}/(\hbar \omega_{\mathrm{c}})$, in order to identify all the possible stable phases of the model. Analytical cross-checks have been carried out whenever possible. 

While our theory is completely general so far, from now on we focus on the resonant regime, i.e.~we set $\epsilon=\hbar \omega_{\mathrm{c}}$.
This clearly enables optimal coupling between the DQDs and the cavity radiation field~\cite{Childress04}. In this case, we have identified four distinct phases: 
\begin{itemize}
\item[i)] A ferromagnetic-normal (FN) phase, with $\mathcal{A}=0$, $s=-1$, and $m=0$. In the language of the original charge degrees of freedom, in the FN phase electrons (holes) occupy the ground state $|{\rm g}\rangle_{e}$ ($|{\rm g}\rangle_{h}$) of each DQD in the top (bottom) chain.
\item[ii)] A ferromagnetic-superradiant (FS) phase, with $\mathcal{A}\neq 0$, $s\neq0$, and $m=0$.
\item[iii)] An antiferromagnetic-normal (AFN) phase, with $\mathcal{A}=0$, $s=0$, and $m=1$. In the language of the original charge degrees of freedom, in the AFN phase electrons occupy the ground state on even sites and the excited state on odd sites and holes in the bottom chain follow the same charge profile.
\item[iv)] An antiferromagnetic-superradiant (AFS) phase, with $\mathcal{A}\neq0$, $s\neq0$, and $m\neq0$. 
\end{itemize}

\begin{figure}[h]
\centering
\begin{overpic}[width=4.45cm]{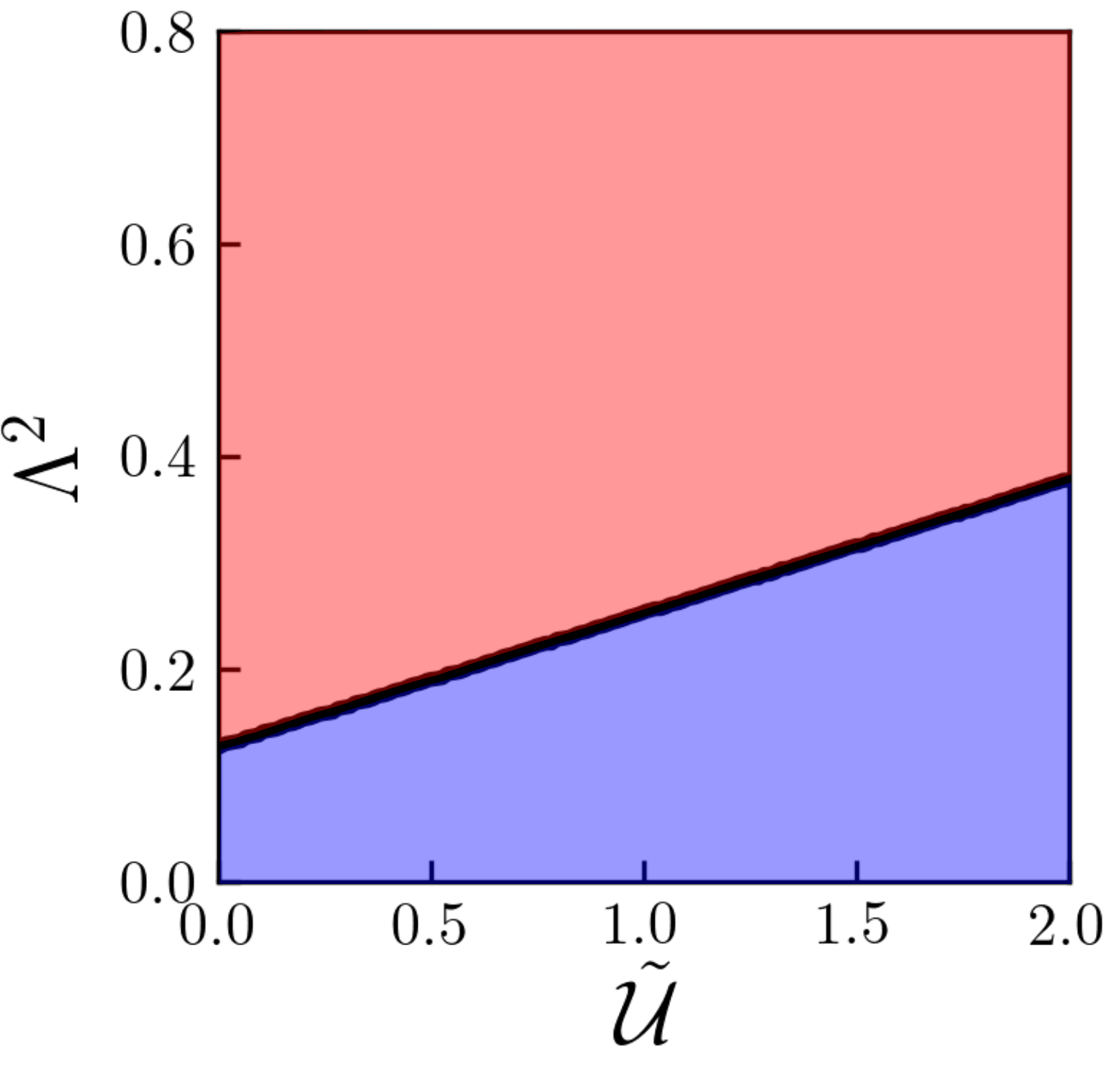}
\put(2, 82){\textbf{a}}
\end{overpic}
\begin{overpic}[width=4.45cm]{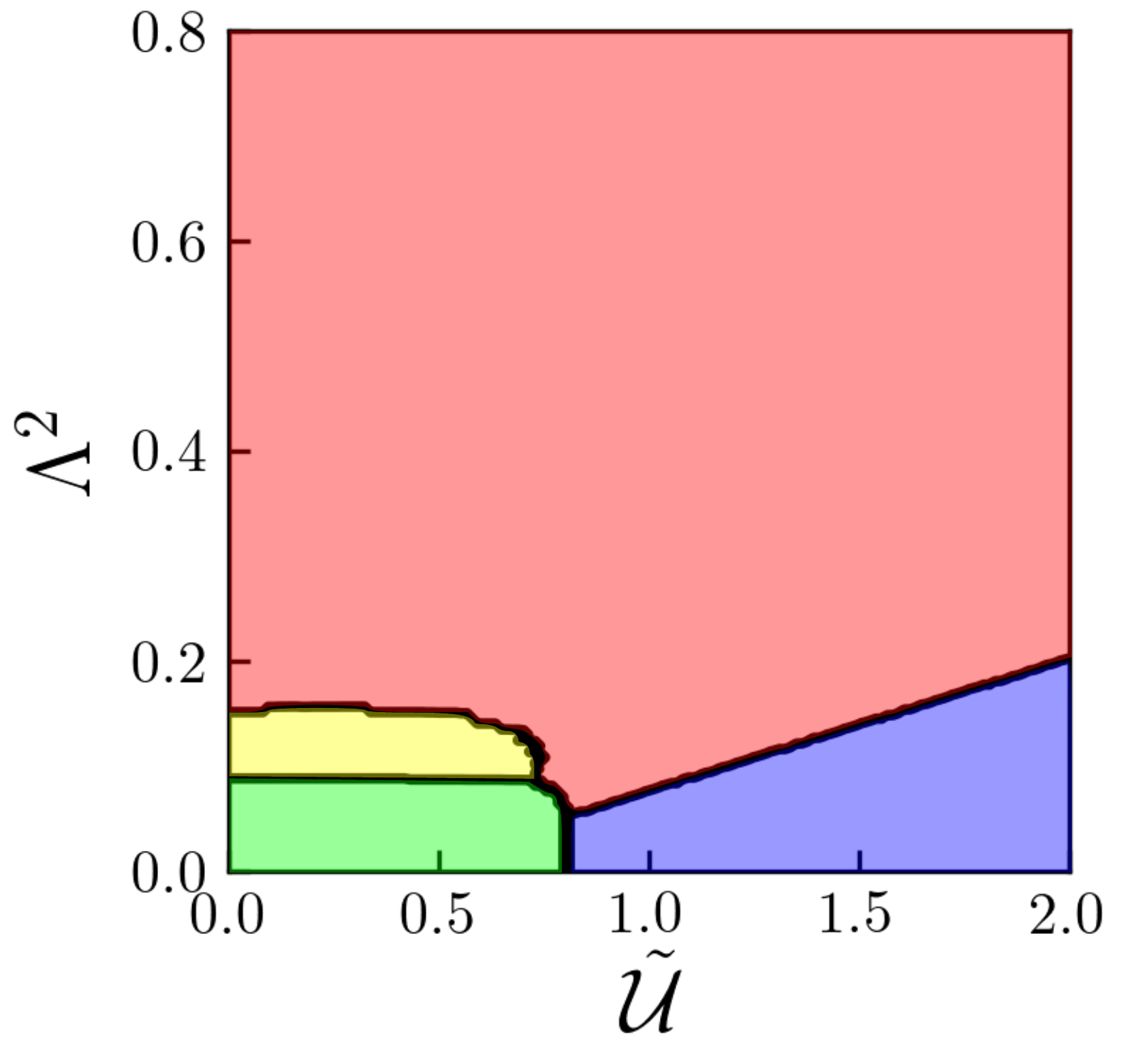}
\put(2, 82){\textbf{b}}
\end{overpic}
\begin{overpic}[width=4.45cm]{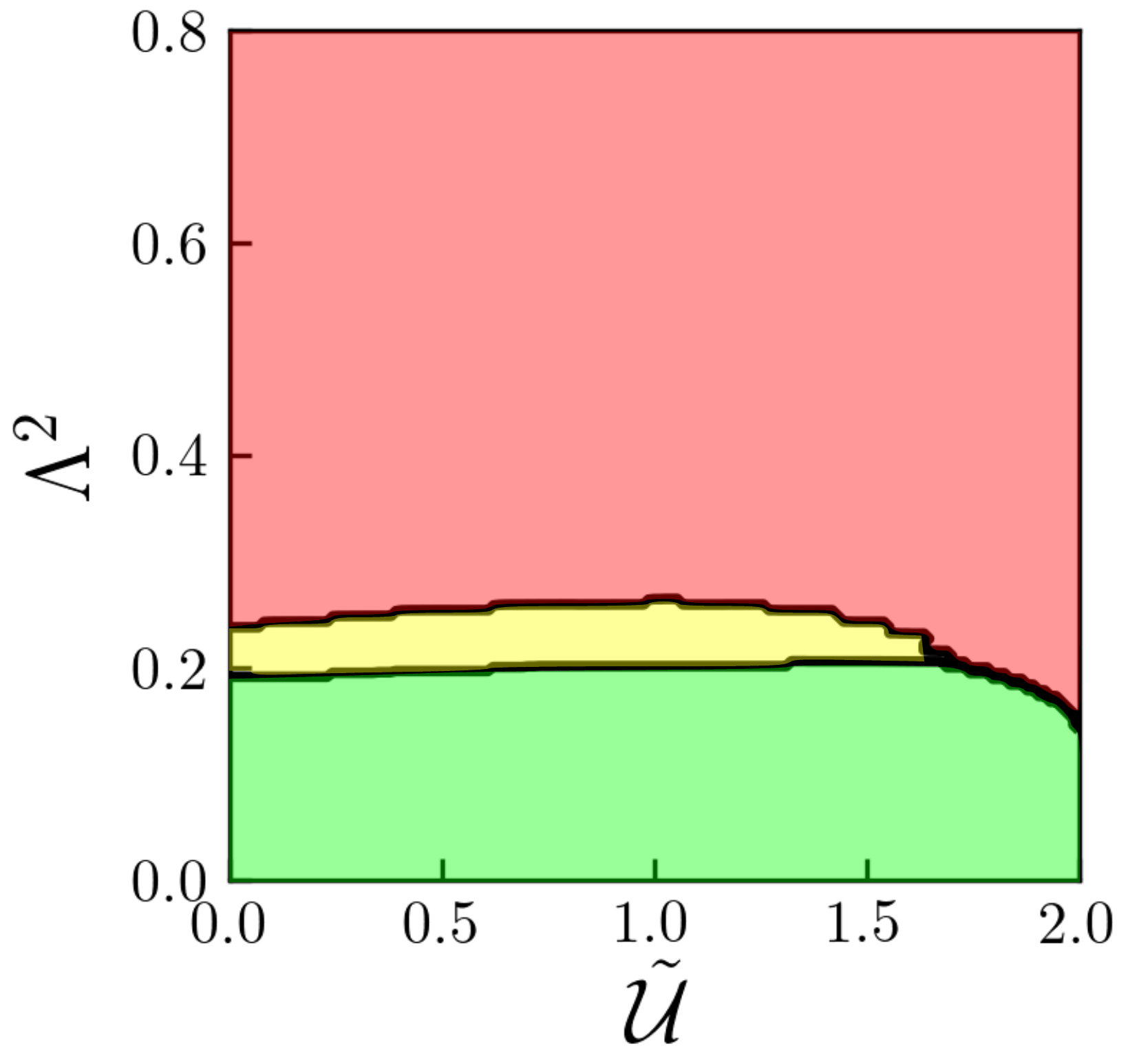}
\put(2, 80){\textbf{c}}
\end{overpic}
\caption{{\textbf{Phases of the quantum supercapacitor model introduced in this work.} (\textbf{a}) At $\tilde{\mathcal{V}}=0$} the phase diagram shows a continuous phase transition between the ferromagnetic-normal (blue) and the ferromagnetic-superradiant (red) ordering, occurring at $\Lambda^{2}=(1+\tilde{\mathcal{U}})/8$. The situation remains qualitatively analogous up to $\tilde{\mathcal{V}} \lesssim0.5$. (\textbf{b}, \textbf{c}) By further increasing $\tilde{\mathcal{V}}$ ($\tilde{\mathcal{V}}=0.7$ in panel {\bf b} and $\tilde{\mathcal{V}}=1.0$ in panel {\bf c}) one observes the emergence of both an antiferromagnetic-normal (green) and a narrow antiferromagnetic-superradiant (yellow) phase at the expense of the previously discussed ones. The ferromagnetic-normal/antiferromagnatic-normal transition is first order and occurs at $\tilde{\mathcal{U}} = 4\tilde{\mathcal{V}}-2$. Notice that this phase diagram has been deduced from the global analysis of the order parameters $\mathcal{A}$, $s$, and $m$, introduced in Eqs.~(\ref{eq:order_parameters}), which are reported in Fig.~\ref{Order_parameters}. The present phase diagram has been calculated for the resonat condition $\epsilon=\hbar \omega_{\rm {c}}$, where the coupling between DQDs and radiation is optimal~\cite{Childress04}.}
\label{Phase_diagram}
\end{figure}

The numerically calculated phase diagram is reported in Fig.~\ref{Phase_diagram} for different values of $\tilde{\mathcal{V}}$. At $\tilde{\mathcal{V}}=0$ (Fig.~\ref{Phase_diagram}a), we observe a net separation between the FN  (blue) and FS phases (red), with a continuous transition occurring at 
\begin{equation}
\Lambda^{2}=\frac{1}{8}\left( 1+\tilde{\mathcal{U}}\right)~.
\end{equation}
With increasing $\tilde{\mathcal{V}}$ (Figs.~\ref{Phase_diagram}b-c), the FN phase shrinks and the AFN phase (green) emerges and expands, extending for small values of $\Lambda^{2}$ up to
\begin{equation}
\tilde{\mathcal{U}} = 4\tilde{\mathcal{V}}-2~,
\end{equation}
at which a first-order transition occurs. Moreover, at the boundary between FS and  AFN phases a very narrow AFS region (yellow) appears. 

\begin{figure}[h]
\centering
\begin{overpic}[width=4.36cm]{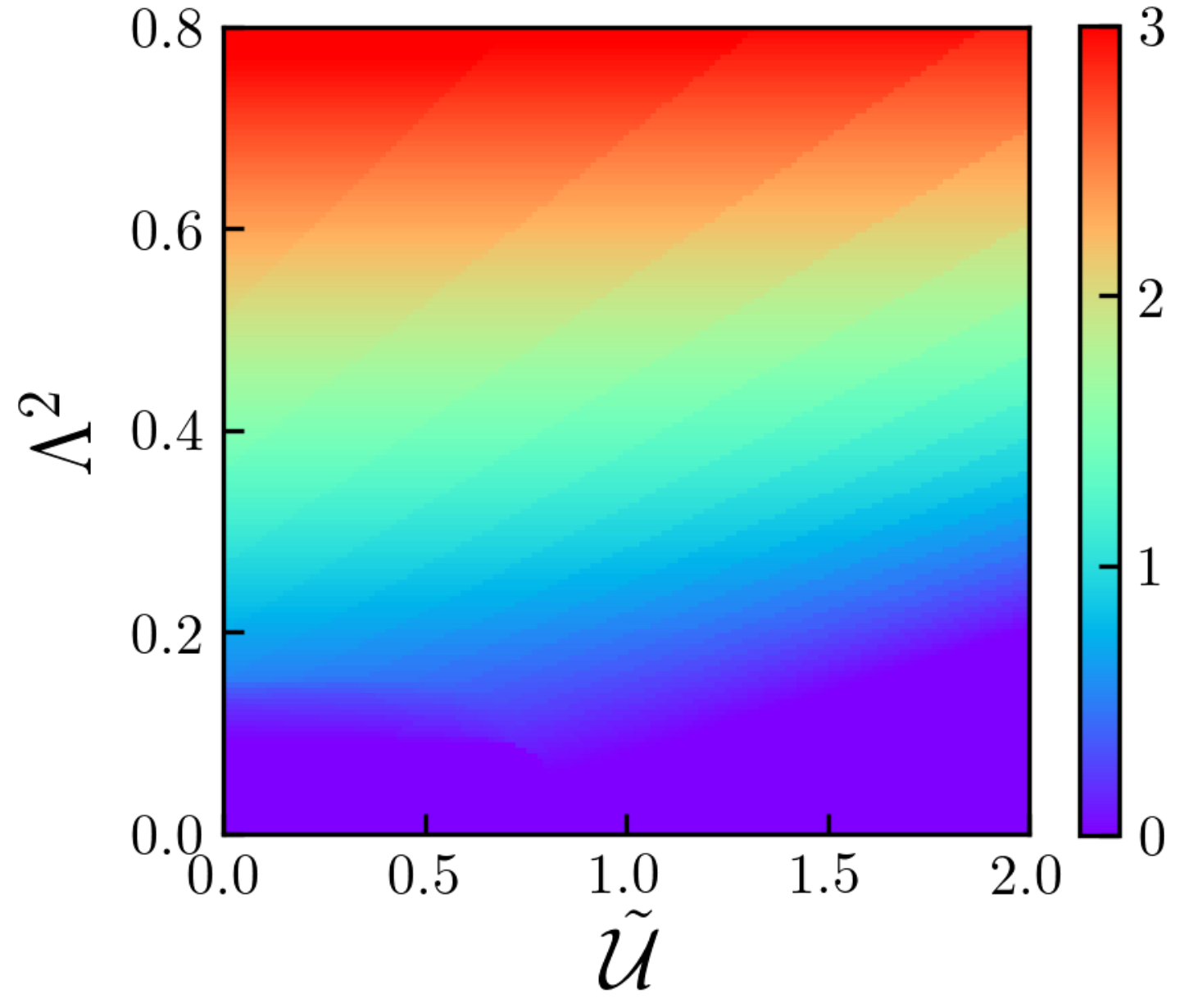}
\put(3, 71){\textbf{a}}
\put(87,84){$\mathcal{A}$}
\end{overpic}
\begin{overpic}[width=4.6cm]{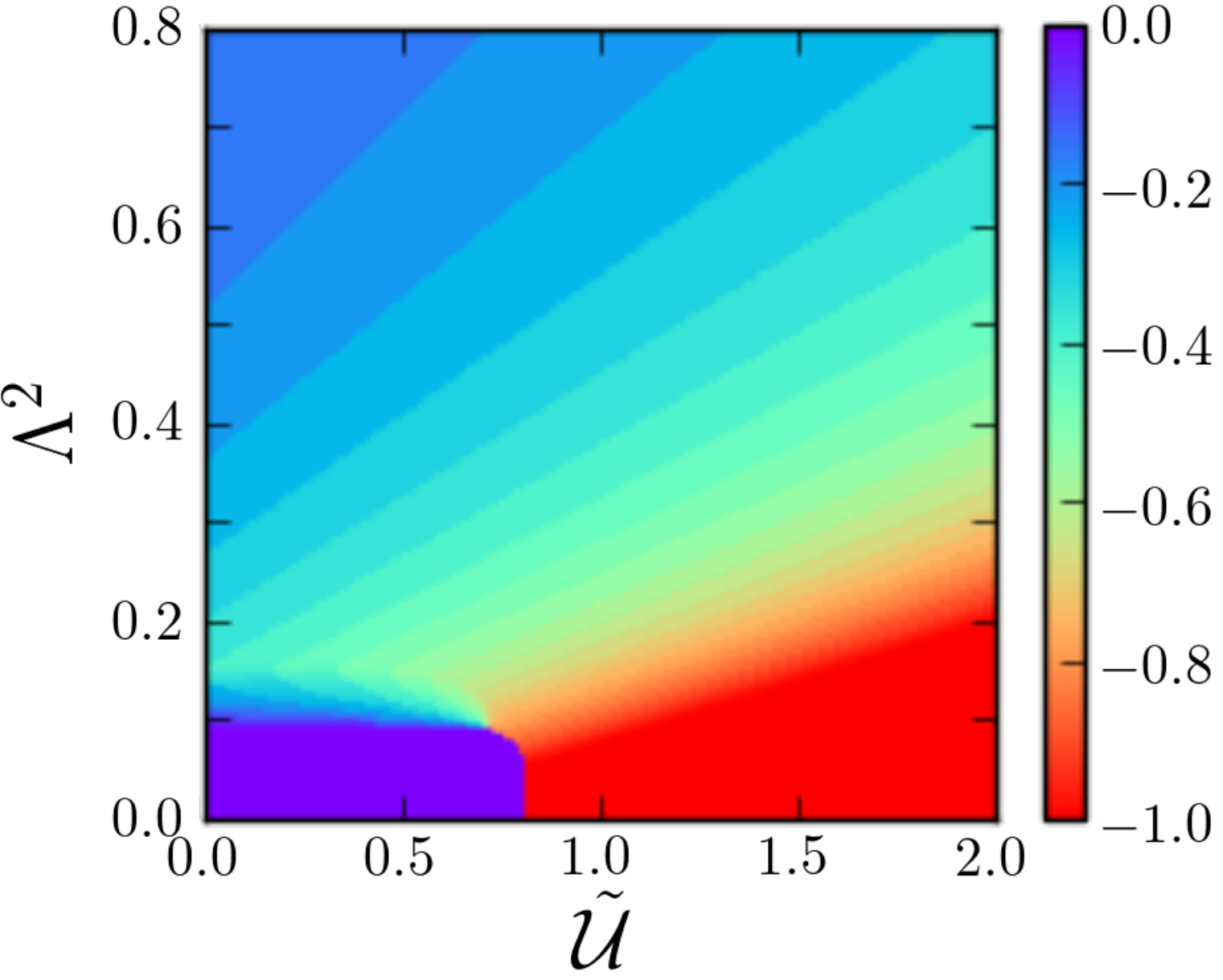}
\put(3, 67){\textbf{b}}
\put(84,79){$s$}
\end{overpic}
\begin{overpic}[width=4.4cm]{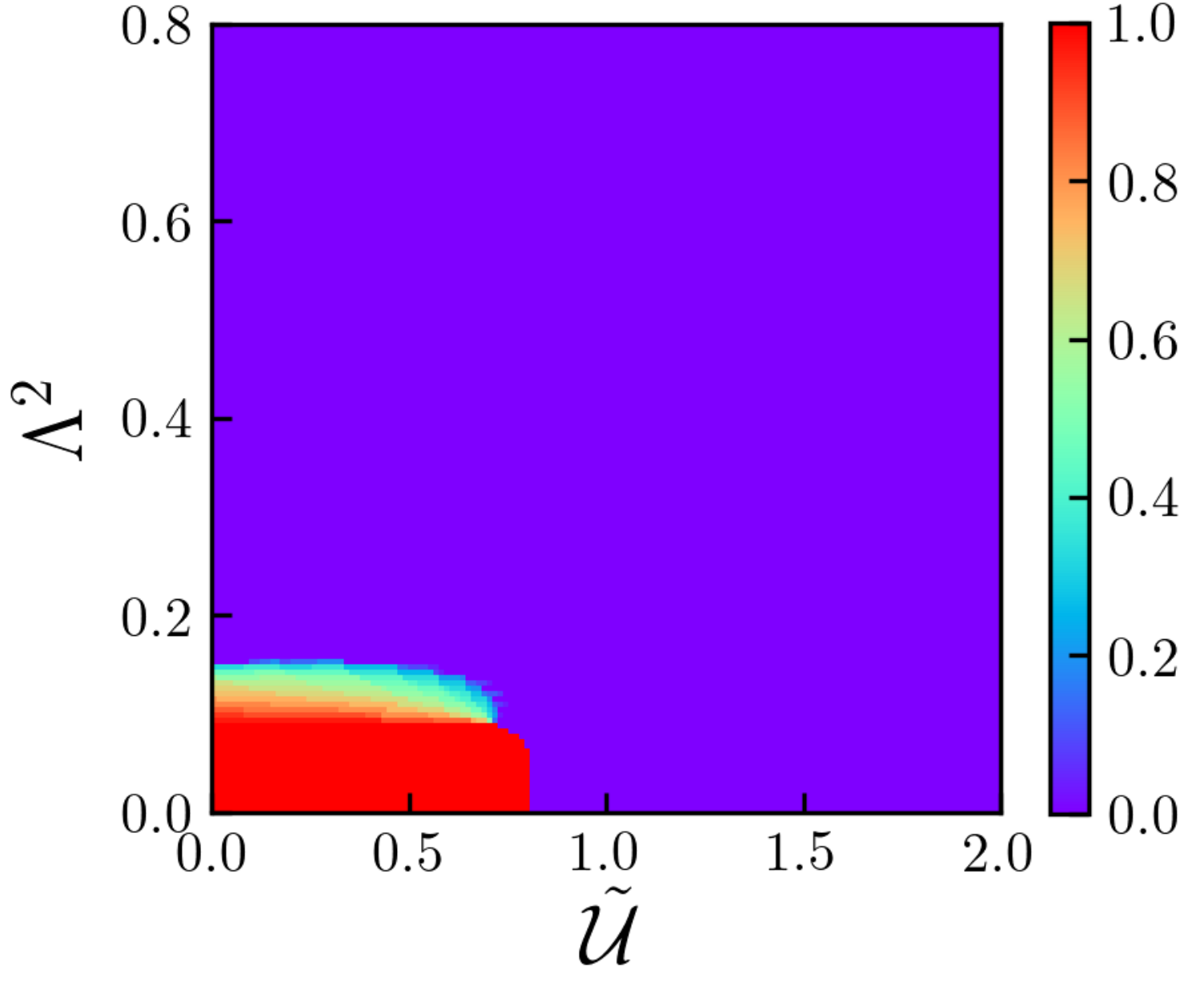}
\put(3, 69){\textbf{c}}
\put(85,82){$m$}
\end{overpic}
\begin{overpic}[width=4.36cm]{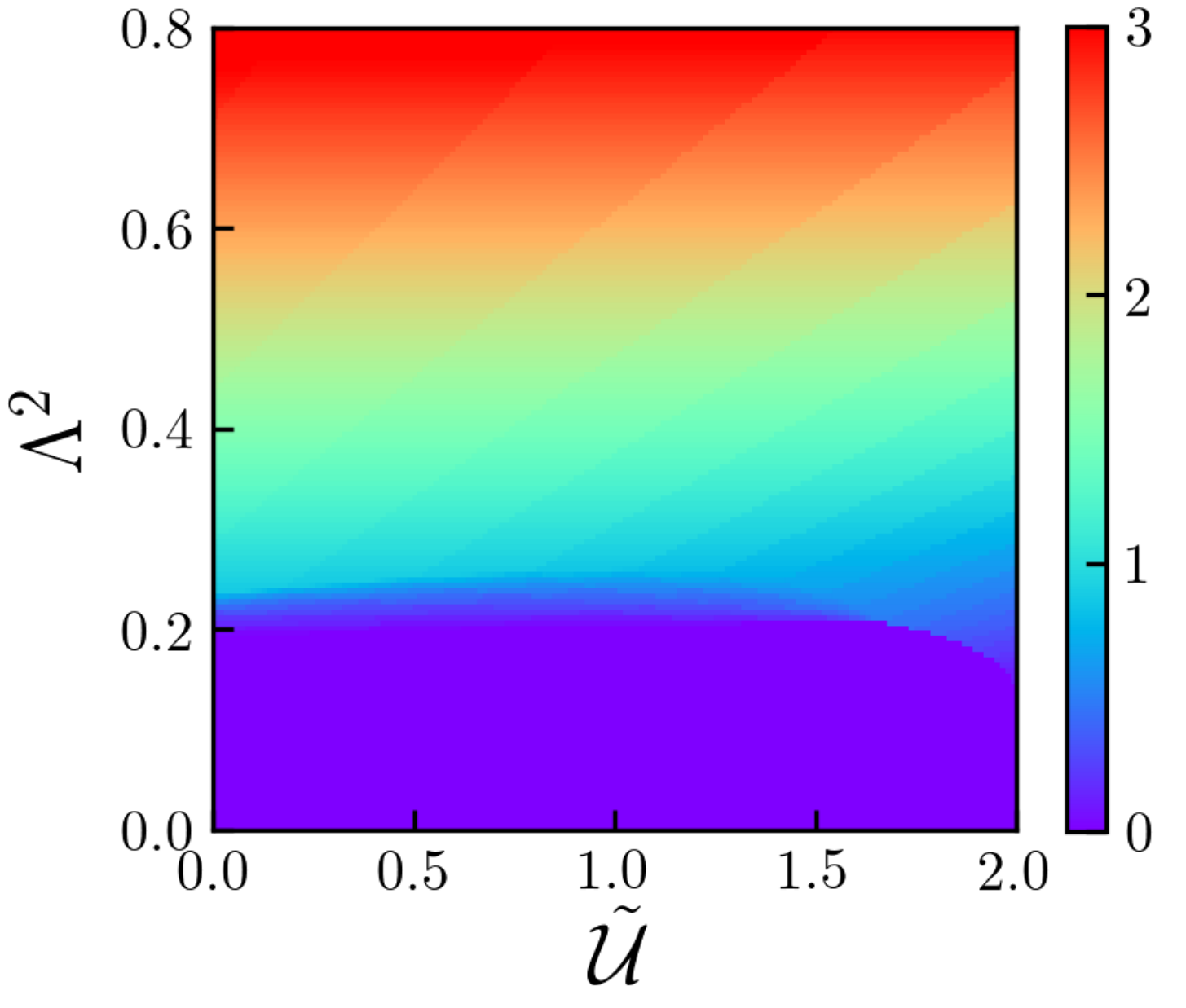}
\put(3, 71){\textbf{d}}
\put(87,85){$\mathcal{A}$}
\end{overpic}
\begin{overpic}[width=4.6cm]{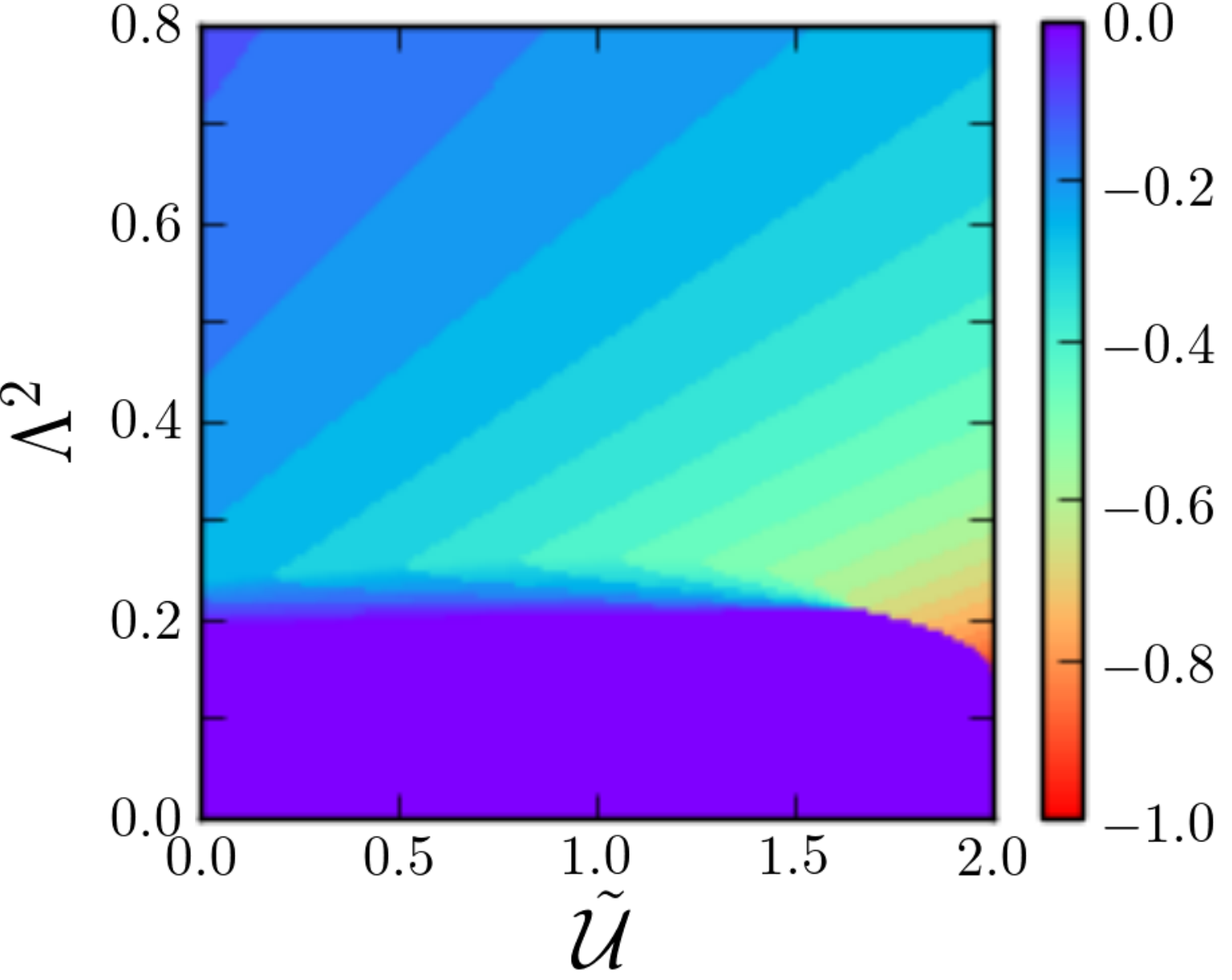}
\put(3, 67){\textbf{e}}
\put(83,79){$s$}
\end{overpic}
\begin{overpic}[width=4.42cm]{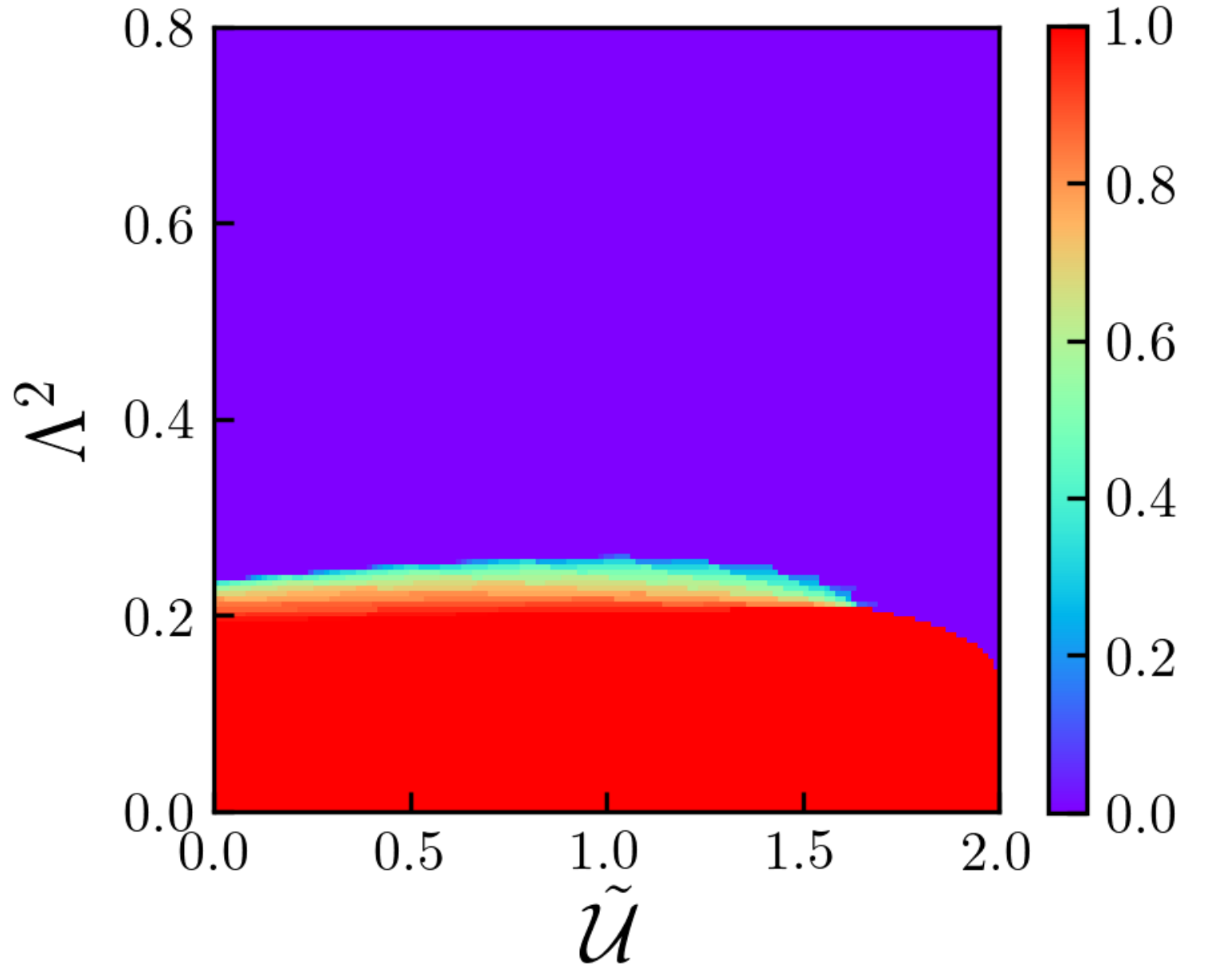}
\put(3, 69){\textbf{f}}
\put(85,81){$m$}
\end{overpic}
\caption{\textbf{Behavior of the order parameters.} Density plots of the order parameters $\mathcal{A}=\langle \hat{a}^{\dagger} \hat{a}\rangle/N$, $s=\langle\left( \hat{\sigma}^{z}_{1}+\hat{\sigma}^{z}_{2}\right)\rangle/2$, and $m=\langle\left( \hat{\sigma}^{z}_{1}-\hat{\sigma}^{z}_{2}\right)\rangle/2$ as functions of $\tilde{\mathcal{U}}$ and $\Lambda^{2}$ for $\tilde{\mathcal{V}}=0.7$ (panels \textbf{a}, \textbf{b}, \textbf{c}) and $\tilde{\mathcal{V}}=1.0$ (\textbf{d}, \textbf{e}, \textbf{f}). Notice that from the top row one can reconstruct the phase diagram of Fig.~\ref{Phase_diagram}b, while the bottom row leads to Fig.~\ref{Phase_diagram}c. Alla data in this figure refer to the resonat condition $\epsilon=\hbar \omega_{\rm {c}}$.}
\label{Order_parameters}
\end{figure}

Our numerical results for the order parameters $\mathcal{A}$, $s$, and $m$ as functions of $\tilde{\mathcal{U}}$ and $\Lambda^{2}$ are reported in the density plots of Fig.~\ref{Order_parameters}. We clearly see that the knowledge of all the three order parameters is needed to properly reconstruct the complete phase diagram. 

{\bf Quantum capacitance.}

We now consider the capacitance for our mesoscopic structure, defined as the inverse of the discrete derivative of the chemical potential with respect to the number $N$ of charges~\cite{Iafrate95, Prus96}, i.e. 
\begin{equation}
C=\frac{e^{2}}{\mu_{N}-\mu_{N-1}}~,
\end{equation}
where $\mu_{N}= E_{N}-E_{N-1}$. Here, $E_{k}$ ($k \in \mathbb N$) indicates the ground-state energy of the system where only $k$ sites of the chains (out of a total of $N$ sites per chain) are filled with electrons and holes. The problem is thereby reduced to evaluating the change in the ground-state energy of the system when an electron and a hole are removed from the same $i$-th site, while keeping fixed the total length of the two coupled chains. This protocol allows to locally preserve the charge neutrality of the system, but explicitly breaks its translational invariance.

According to this, it is possible to consider the following two-step protocol. One can first remove an electron-hole pair from the completely filled two-chain system in an arbitrary site (in the sublattice of odd sites, say). Considering the $N\gg 1$ limit, one has
\begin{equation}
\mu_N\approx \epsilon\cos(\theta_{\rm o}) -\frac{\mathcal{U}}{4} \left[1-\cos(\theta_{\rm o})\right]^{2}+2 \mathcal{V} \left[ \cos(\theta_{\rm o}) \cos(\theta_{\rm e})+1\right] -8 \hbar \omega_{\mathrm{c}} \Lambda^{2} \sin^{2}(\theta_{\rm o})~.
\end{equation}
 The second electron-hole pair can be removed in one of the nearest-neighbor sites (both obviously in the sublattice of even sites) leading to
\begin{equation}
\mu_{N-1}\approx \epsilon\cos(\theta_{\rm e})-\frac{\mathcal{U}}{4} \left[1-\cos(\theta_{\rm e})\right]^{2}+\mathcal{V} \left[\cos(\theta_{\rm o}) \cos(\theta_{\rm e})+1\right] -8 \hbar \omega_{\mathrm{c}} \Lambda^{2}\sin^{2}(\theta_{\rm e}).
\end{equation}
The above protocol reminds of what happens in atomic physics, where an atom with a completely filled shell is progressively ionized by removing the most loosely bound electrons and with $\mu_{N}$ ($\mu_{N-1}$) playing the role of first (second) ionization energy~\cite{Iafrate95}.

Notice that corrections scaling as $1/N$ can be taken into account, but have a negligible effect in the behavior of the capacitance.

The capacitance of the system can then be written, considering again the resonant condition  $\epsilon=\hbar \omega_{\mathrm{c}}$, as 
\begin{equation}\label{eq:simple_formula}
C= \frac{C_{0}}{\big[ (2+\tilde{\mathcal{U}}) m-\tilde{\mathcal{U}} m s+\tilde{\mathcal{V}}\left(1+ s^{2}-m^{2} \right)+32 \Lambda^{2} m s \big]}
\end{equation}
with $C_{0} \equiv e^{2}/\hbar \omega_{\mathrm{c}}$ a dimensional factor.

Considering the four physically relevant ground-state phases discussed above, Eq.~(\ref{eq:simple_formula}) shows that, in absence of coupling with the cavity radiation ($\Lambda^{2}=0$), one has $C_{\mathrm{FN}}=C_{0}/2 \tilde{\mathcal{V}}$ for the FN phase ($s=-1, m=0$) and $C_{\mathrm{AFN}}= C_{0}/(2+\tilde{\mathcal{U}})$ (lower than $C_{\mathrm{FN}}$ in the considered range of parameters) for the AFN ($s=0, m=1$) phase. At finite values of the light-matter coupling, in the FS ($s\neq 0, m=0$) phase, the capacitance becomes $C_{\mathrm{FS}}=C_{0}/(1+s^{2}) \tilde{\mathcal{V}}$.

\begin{figure}[h]
\centering
\begin{overpic}[width=6.7cm]{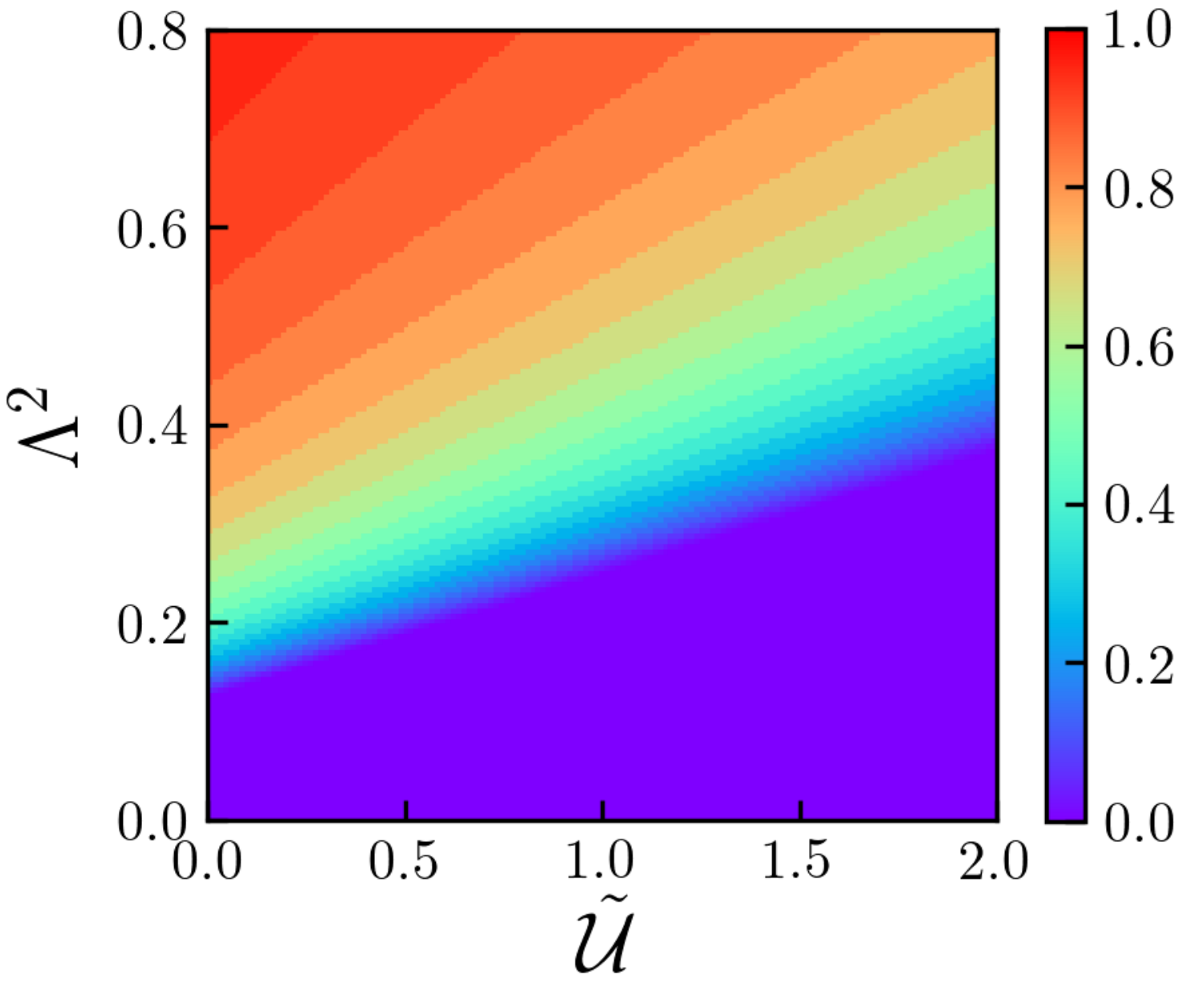}
\put(3, 69){\textbf{a}}
\put(87,82){$\kappa$}
\end{overpic}
\begin{overpic}[width=6.7cm]{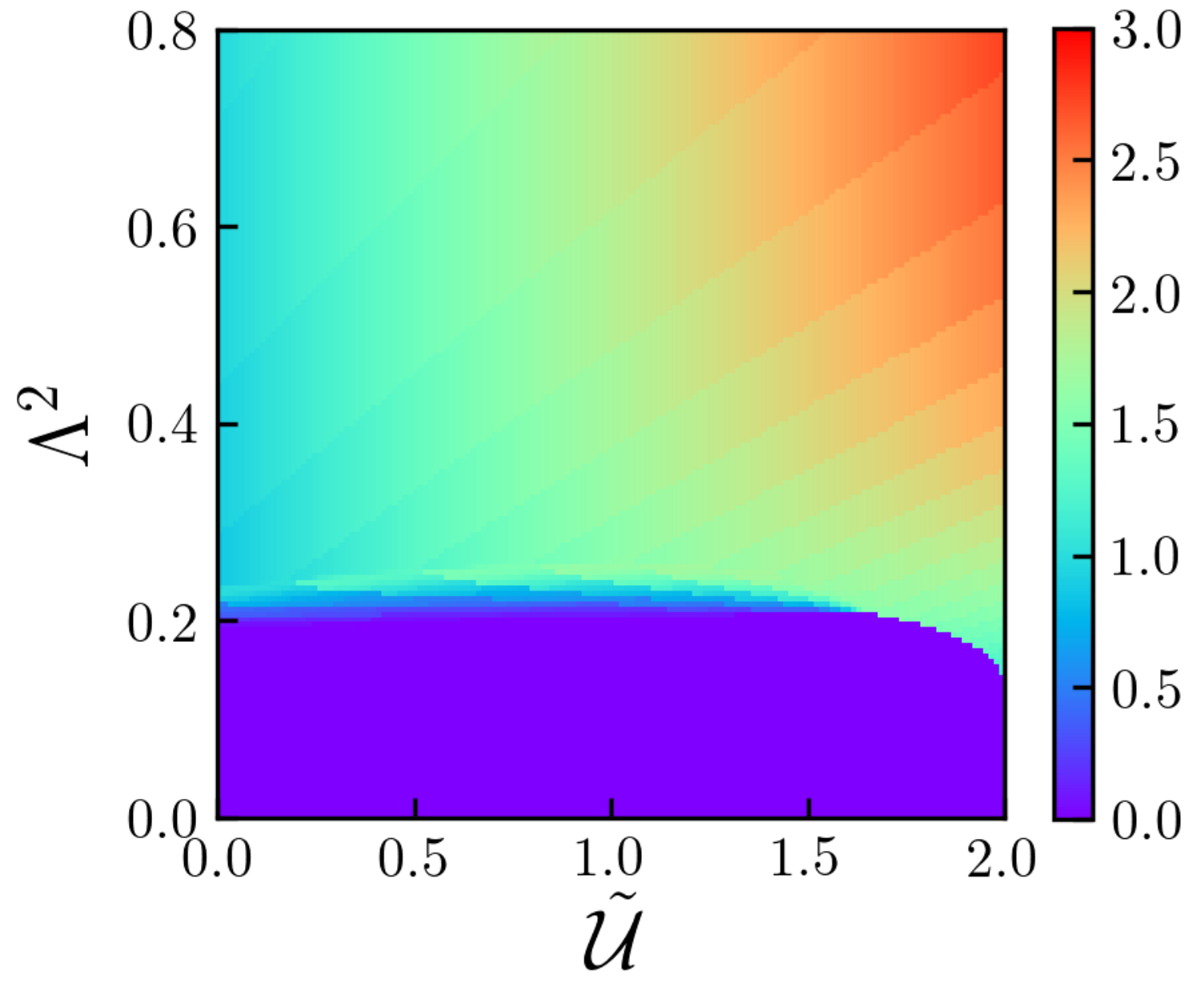}
\put(3, 69){\textbf{b}}
\put(87,82){$\kappa$}
\end{overpic}
\caption{\textbf{Enhancement of the quantum capacitance.} Density plots
of the ratio  $\kappa= C/\bar{C}-1$ (with $C$ the capacitance of the system and $\bar{C}$ its value at $\Lambda^{2}=0$) as a function of $\tilde{\mathcal{U}}$ and $\Lambda^{2}$ for: (\textbf{a}) $\tilde{\mathcal{V}} = 0$ (corresponding to the phase diagram in Fig.~\ref{Phase_diagram}a) where, in the ferromagnetic-superradiant phase, one has $\kappa=(1-s^{2})/(1+s^{2})$. This ratio asymptotically approaches $\kappa=1$ (doubling of the capacitance) at high values of $\Lambda^{2}$. (\textbf{b}) $\tilde{\mathcal{V}} = 1.0$ (corresponding to the phase diagram in Fig.~\ref{Phase_diagram}c) where $\kappa=  (2+\tilde{\mathcal{U}})/\left(1+s^{2}\right)\tilde{\mathcal{V}}-1$ crucially depends on the geometry of the device. Here, for the considered values of the parameters, the ratio can exceed $\kappa >2.5$ (red region) leading to a very remarkable enhancement of the capacitance with respect to the reference case in absence of radiation.}
\label{kappa}
\end{figure}

In this work we are interested in quantifying the enhancement of the capacitance with respect to the one evaluated in absence of radiation, i.e.~$C$ evaluated at $\Lambda^2=0$ and indicated with $\bar{C}$ in the following. To this end we introduce the ratio 
\begin{equation}
\kappa= \frac{C-\bar{C}}{\bar{C}}~. 
\end{equation}
Density plots of this ratio as a function of $\tilde{\mathcal{U}}$ and $\Lambda^{2}$, for different values of $\tilde{\mathcal{V}}$ are reported in Fig.~\ref{kappa}. We note that the ratio is positive and that a strong enhancement of the capacitance ($\kappa >0$) associated to the superradiant phase transition occurrs in the system. In particular, at the transition between the FN ($s=-1, m=0$) and FS ($s\neq 0, m=0$) phase, we have
\begin{equation}
\kappa=\frac{C_{\mathrm{FS}}}{C_{\mathrm{FN}}}-1= \frac{1-s^{2}}{1+s^{2}}~,
\end{equation}
as shown in Fig.~\ref{kappa}a. This quantity only depends on the pseudospin order parameter $s$ and saturates to $\kappa=1$ (i.e.~doubling of the capacitance) deeply in the FS phase (where one asymptotically approaches $s=0$). Importantly, this quantity depends on $1-s^{2}$, which is non-zero only when the ground-state wave-function displays macroscopic quantum coherence. In this case, the appearance of coherence, a genuinely quantum effect, can strongly enhance the capacitance. Differently, at the transition between the AFN ($s=0, m=1$) and  FS ($s\neq 0, m=0$) phase and neglecting the small AFS phase, we find
\begin{equation}
\kappa= \frac{C_{\mathrm{FS}}}{C_{\mathrm{AFN}}}-1= \frac{2+\tilde{\mathcal{U}}}{\left(1+s^{2}\right)\tilde{\mathcal{V}}}-1~,
\end{equation}
which depends on both the pseudospin order parameter $s$ and the specific values of the interaction terms (i.e.~geometry of the device) and can exceed $250\%$ ($\kappa>2.5$), as shown in Fig.~\ref{kappa}b. Again, this quantity is maximal when $s=0$, which corresponds to the fact that the ground-state wave-function is in a equal macroscopic superposition of $|{\rm{g}}\rangle$ and $|{\rm{e}}\rangle$.
As a matter of fact, radiation induces macroscopic coherence, which, in turn, strongly enhances the capacity.

{\bf Discussion.} 

Before concluding, we comment on the possibility of realistic solid state implementations of our quantum supercapacitor model in terms of coupled charge qubits (DQDs) embedded  in a microwave cavity.

Recent investigations of two-qubit logic gates made up of GaAs/AlGaAs and Si/SiGe DQDs~\cite{Ward16} have reported capacitive coupling between DQDs up to $\mathcal{U}/ h\approx 30~{\rm GHz}$. Assuming level spacings and cavity frequencies $\epsilon /h\approx \omega_{\mathrm{c}}/2 \pi$ in the GHz range as typical of mesocopic devices~\cite{Frey12, Toida13, Fink09}, we conclude that it is possible, at least in principle, to explore a quite wide interval of values of $\tilde{\mathcal{U}}$. Moreover, one can also change this parameter by both acting on the distance $d_{\perp}$ between the two chains and changing the dielectric constant $\varepsilon$ of the environment where the chains are embedded according to the relation
\begin{equation}
\mathcal{U}= \frac{e^{2}}{\varepsilon d_{\perp}}~.
\end{equation}
An analogous discussion also holds for the intra-chain coupling,
\begin{equation}
\mathcal{V}= \frac{e^{2}}{\varepsilon d_{\parallel}}~,
\end{equation} 
where $d_{\parallel}$ is the distance between the DQDs along each chain. 

Concerning the coupling between the DQDs and the cavity radiation, experimental techniques allowing to explore interactions up to $\Lambda^{2} \approx 1$, deep in the superradiant phase, can be envisaged~\cite{Frey12, Toida13, Yoshihara17, Langford17, Braumuller17}. Despite this, the actual possibility to explore the normal/superradiant phase transition in a real
solid-state device has been debated at length~\cite{Chen07, Nataf10, Viehmann11, Chirolli12, Pellegrino14, Pellegrino16} due to the presence of an additional term $\propto \left(\hat{a}^{\dagger}+\hat{a} \right)^{2}$---not included in the simple model Hamiltonian in Eq.~(\ref{Starting_H})---which emerges from the minimal coupling between matter and cavity radiation.  However, according to recent calculations~\cite{Mazza19}, superradiance can occur in correlated materials embedded in photonic cavities.

As a final overall experimental requirement one need to consider devices with a typical decoherence ($\tau_{\varphi}$) and relaxation ($\tau_{\rm{r}}$) times long enough to be effectively operated \cite{Blais04, Schoelkopf08}. This condition is satisfied for example in the experiment discussed in Ref.~\cite{Stockklauser17} and is a common constraint for all quantum devices.

Accordingly, a realization of the proposed device with supercapacitive  properties is feasible, e.g.~by means of circuit QED devices, comprising two chains of charge qubits and a resonator.

{\bf Methods.}

{\bf Mapping between electrons and spins and link between electrons and holes.} We here derive the pseudospin model introduced in Eq.~(\ref{Starting_H}) starting from a fermionic model of two coupled (top and bottom) chains of electrons and holes.
We denote by $e_{{\rm g},i},e^\dagger_{{\rm g},i}$ ($e_{{\rm e},i},e^\dagger_{{\rm e},i}$) the fermionic annihiliation and creation ladder operators for an electron with onsite energy  $\epsilon_{\rm g}$ ($\epsilon_{\rm e}$) residing on the $i$-th site of a chain. It is useful to define the onsite energy difference $\epsilon=\epsilon_{\rm e}-\epsilon_{\rm g}$. The electronic Hamiltonian (for the top chain) reads:
\begin{eqnarray}
\label{Starting_Hel}
 \hat{\mathcal{H}}^{({\rm T})}_{\rm DI}&=&\sum_i\Big[    \epsilon_{\rm g} e^\dagger_{{\rm g},i}e_{{\rm g},i}    + \epsilon_{\rm e} e^\dagger_{{\rm e},i}e_{{\rm e},i}    \nonumber \\& +&    \mathcal{V}\Big( e^\dagger_{{\rm g},i}e_{{\rm g},i} e^\dagger_{{\rm g},i+1}e_{{\rm g},i+1} +  e^\dagger_{{\rm e},i}e_{{\rm e},i} e^\dagger_{{\rm e},i+1}e_{{\rm e},i+1}    \Big)  
\nonumber \\& +& \lambda\hbar\omega_{\mathrm c} (a+a^\dagger)\Big( e^\dagger_{{\rm g},i}e_{{\rm e},i}    + e^\dagger_{{\rm e},i}e_{{\rm g},i}     \Big)             \Big]~,
\end{eqnarray}
where $\lambda$, $\mathcal{V}$, $\omega_{\rm c}$, $a$, and $a^\dagger$ have been introduced in the main text. In order to introduce the pseudospin representation, we write the ladder operators in a natural basis. We denote by $\ket{k,l}_{{\rm e},i}$ the state with $k=0,1$ ($l=0,1$) electrons in the local ground (excited) state on the $i$-th site.
 We assume to have $1$ electron per site. (All the interactions in the Hamiltonian preserve the local number of electrons $\hat{n}_{{\rm e},i}=  e^\dagger_{{\rm g},i}e_{{\rm g},i} +e^\dagger_{{\rm e},i}e_{{\rm e},i}$.) Hence we can expand  Eq.~(\ref{Starting_Hel}) in the basis $\big\{\ket{0,1}_{{\rm e},i},\ket{1,0}_{{\rm e},i}\big\}$, which means that local operators on the $i$-th site admits an Hermitian $2 \times 2$ representation, which can be written in terms of Pauli matrices. In this basis, the Hamiltonian in Eq.~(\ref{Starting_Hel}) reads
\begin{equation}\label{eq:DIT2}
\hat{\mathcal{H}}^{({\rm T})}_{\rm DI}=\frac{\epsilon}{2} \sum_{i=1}^{N} \hat{\tau}^{z}_{i}+\frac{\mathcal{V}}{2} \sum_{i=1}^{N} \left( \hat{\tau}^{z}_{i} \hat{\tau}^{z}_{i+1}+1\right) + \hbar \omega_{\mathrm c} \lambda \left(\hat{a}^{\dagger}+\hat{a} \right)\sum_{i=1}^{N} \hat{\tau}^{x}_{i}~,
\end{equation} 
which is exactly the form used in the main text.

We now move on to analyze the bottom chain, which hosts holes. We start from the simple observation that hole annihilation and creation ladder operators, $h$ and $h^\dagger$, can be obtained from those for electrons, $e$ and $e^\dagger$, by using a particle-hole transformation: $e\to h^\dagger$ and $e^\dagger \to h$. Starting from Eq.~(\ref{Starting_Hel}) and carrying out this transformation, we can write the Hamiltonian of the bottom chain as
\begin{eqnarray}
\label{Starting_Hhole}
 \hat{\mathcal{H}}^{({\rm B})}_{\rm DI}&=&\sum_i\Big[  -  \epsilon_{\rm g} h^\dagger_{{\rm g},i}h_{{\rm g},i}    - \epsilon_{\rm e} h^\dagger_{{\rm e},i}h_{{\rm e},i}     \nonumber \\
 &+&   \mathcal{V}\Big( h^\dagger_{{\rm g},i}h_{{\rm g},i} h^\dagger_{{\rm g},i+1}h_{{\rm g},i+1} +  h^\dagger_{{\rm e},i}h_{{\rm e},i} h^\dagger_{{\rm e},i+1}h_{{\rm e},i+1}    \Big)   \nonumber \\
 &+&\lambda\hbar\omega_{\mathrm c} (a+a^\dagger)\Big( h^\dagger_{{\rm g},i}h_{{\rm e},i}    + h^\dagger_{{\rm e},i}h_{{\rm g},i}\Big)\Big]~,
\end{eqnarray}
where $h_{{\rm g},i},h^\dagger_{{\rm g},i}$ ($h_{{\rm e},i},h^\dagger_{{\rm e},i}$) are ladder operators for a hole with onsite energy  $-\epsilon_{\rm g}$ ($-\epsilon_{\rm e}$) residing on the $i$-th site.

It is useful to denote by $\ket{k,l}_{{\rm h},i}$ the state with $k=0,1$ ($l=0,1$) holes in the local ground (excited) state on the $i$-th site.
Again we assume to have 1 hole per site, since all the interactions in the Hamiltonian preserve the local number of holes. Expanding Eq.~(\ref{Starting_Hhole}) in the basis $\big\{\ket{1,0}_{{\rm h},i},\ket{0,1}_{{\rm h},i}\big\}$, we get

\begin{equation}\label{eq:DIT3}
\hat{\mathcal{H}}^{({\rm B})}_{\rm DI}=\frac{\epsilon}{2} \sum_{i=1}^{N} \hat{\sigma}^{z}_{i}+\frac{\mathcal{V}}{2} \sum_{i=1}^{N} \left( \hat{\sigma}^{z}_{i} \hat{\sigma}^{z}_{i+1}+1\right) + \hbar \omega_{\mathrm c} \lambda \left(\hat{a}^{\dagger}+\hat{a} \right)\sum_{i=1}^{N} \hat{\sigma}^{x}_{i}~,
\end{equation} 
which is exactly the form used in the main text.
Finally we analyze the term~\cite{Hubbard63}
\begin{equation}\label{eq:intra_chainEL}
\hat{\mathcal{H}}^{({\rm TB})} = -\mathcal{U} \sum_{i=1}^{N}  c^\dagger_{{\rm g},i}c_{{\rm g},i} h^\dagger_{{\rm e},i}h_{{\rm e},i}~,
\end{equation}
which represents a local attractive interaction between electrons and holes in the two adjacent chains. Expanding this Hamiltonian in the aforementioned basis, we immediately get Eq.~(\ref{eq:intra_chain}). We would finally like to make contact with the notation used in Fig.~\ref{fig1}. In the latter, we defined $\ket{g}=\ket{1,0}_{{\rm e},i}$ and $\ket{e}=\ket{0,1}_{{\rm e},i}$ for the top chain, and $\ket{g}=\ket{0,1}_{{\rm h},i}$ and $\ket{e}=\ket{1,0}_{{\rm h},i}$ for the bottom chain.

\newpage

\end{document}